\documentclass[iop,revtex4]{emulateapj}

\usepackage{hyperref} 

\begin{document}
\slugcomment{Accepted for publication in ApJ, May 30, 2014}

\shorttitle{Doppler Tomography of Kepler-13 A\lowercase{b}}
\shortauthors{Johnson et al.}

\title{A Misaligned Prograde Orbit for Kepler-13\,A\lowercase{b} via Doppler Tomography\altaffilmark{$\star$}}

\author{Marshall C. Johnson\altaffilmark{1}, William D. Cochran\altaffilmark{1},  Simon Albrecht\altaffilmark{2}, Sarah E. Dodson-Robinson\altaffilmark{3}, Joshua N. Winn\altaffilmark{2}, and Kevin Gullikson\altaffilmark{1}}

\altaffiltext{1}{Department of Astronomy and McDonald Observatory, University of Texas at Austin, 2515 Speedway, Stop C1400, Austin, TX 78712, USA; mjohnson@astro.as.utexas.edu}
\altaffiltext{2}{Department of Physics and Kavli Institute for Astrophysics and Space Research, Massachusetts Institute of Technology, Cambridge, MA 02139, USA}
\altaffiltext{3}{Department of Physics and Astronomy, University of Delaware, 217 Sharp Lab, Newark, DE 19716, USA}
\altaffiltext{$\star$}{Based in part on observations obtained with the Hobby-Eberly Telescope, which is a joint project of the University of Texas at Austin, the Pennsylvania State University, Stanford University, Ludwig-Maximilians-Universit\"at M\"unchen, and Georg-August-Universit\"at G\"ottingen}

\begin{abstract}

Transiting planets around rapidly rotating stars are not amenable to precise radial velocity observations, such as are used for planet candidate validation, as they have wide, rotationally broadened stellar lines. Such planets can, however, be observed using Doppler tomography, wherein the stellar absorption line profile distortions during transit are spectroscopically resolved. This allows the validation of transiting planet candidates and the measurement of the stellar spin-planetary orbit (mis)alignment, an important statistical probe of planetary migration processes. We present Doppler tomographic observations which provide a direct confirmation of the hot Jupiter Kepler-13 Ab, and also show that the planet has a prograde, misaligned orbit, with $\lambda=58.6^{\circ} \pm 2.0^{\circ}$. Our measured value of the spin-orbit misalignment is in significant disagreement with the value of $\lambda=23^{\circ} \pm 4^{\circ}$ previously measured by \cite{Barnes11} from the gravity-darkened {\it Kepler} lightcurve. 
We also place an upper limit of $0.75 M_{\odot}$ (95\% confidence) on the mass of Kepler-13 C, the spectroscopic companion to Kepler-13 B, the proper motion companion of the planet host star Kepler-13 A.
\end{abstract}


\section{Introduction}
\label{introduction}

Observations over the past few years have shown that many transiting exoplanets (principally hot Jupiters) have significantly non-zero orbital inclinations. This is, in most cases, measured via the Rossiter-McLaughlin effect, which is a probe of the sky-projected orbital inclination ($\lambda$), also known as the spin-orbit misalignment \citep[e.g.,][]{Triaud10,Winn11}. \cite{Winn10} noted two different regimes in the distribution of $\lambda$ versus stellar $T_{\mathrm{eff}}$. Planets orbiting cooler stars ($T_{\mathrm{eff}}<6250$~K) tend to have aligned orbits (with a few notable exceptions), while those orbiting hotter stars ($T_{\mathrm{eff}}>6250$ K) have a much wider distribution of spin-orbit misalignments that is consistent with isotropic \citep{Albrecht12}.

Several hypotheses have been put forward to explain these two regimes. \cite{Winn10} proposed that most hot Jupiters are emplaced on highly inclined orbits by processes such as planet-planet scattering \citep[e.g.,][]{RasioFord96} or Kozai cycles \citep[e.g.,][]{KCTF}. $T_{\mathrm{eff}}=6250$ K marks the location on the main sequence where cooler stars have deep, massive convective zones, whereas hotter stars do not. \cite{Winn10} hypothesized that cooler stars' convective zones are able to efficiently tidally couple to the planet and damp out the planetary orbital inclination within the main sequence lifetime, whereas those of hotter stars are not. \cite{ValsecchiRasio14} recently presented simulations confirming the plausibility of this idea. \cite{Batygin12} instead proposed that hot Jupiters are emplaced by disk migration within an inclined disk, coupled with the same tidal dissipation hypothesis as \cite{Winn10}. The disk is torqued out of alignment with the stellar spin axis by gravitational interactions with a transitory binary companion on an inclined orbit in the birth cluster. Further simulations along these same lines by \cite{BatyginAdams13} and \cite{Lai14} included magnetic and gravitational interactions between the host star and the disk; both found that this remains a viable misalignment mechanism. \cite{Lai11} had earlier found that magnetic interactions between the star and disk alone could torque the star out of alignment with the disk. \cite{Bate10} argued that time variability in the bulk angular momentum of material being accreted by a protoplanetary system could result in a spin-orbit misalignment between the star and the planet-forming disk. Another mechanism was proposed by \cite{Rogers12}, who modeled angular momentum transport via internal gravity waves within hot stars, and suggested that such angular momentum transport could drastically change the rotational properties of the stellar atmosphere on short time scales. The rotation of the stellar atmosphere, which is what is probed by all spin-orbit misalignment measurement techniques, would not reflect the bulk rotation of the star. An apparent spin-orbit misalignment could thus be generated even when the bulk angular momentum vectors of the star and planet are in fact well aligned. Furthermore, \cite{Rogers13} called into question whether tidal damping could affect inclinations as proposed by \cite{Winn10} and \cite{Batygin12}. \cite{Rogers13} found that in order for tidal damping not to result in significant semi-major axis changes, inclinations must be driven to $0^{\circ}, \pm 90^{\circ}, 180^{\circ}$, which is not observed. \cite{Xue14}, however, showed that the latter two of these states would eventually decay to the zero inclination state. In general, these hypotheses fall into two categories: either the planets have changed their orbital plane after their formation, or the planetary orbit and stellar rotation axes are misaligned for reasons unrelated to planet evolution, and are related to star formation or stellar physics. Measurement of the spin-orbit misalignments of a statistically significant sample of long-period planets (which should not have undergone significant tidal damping) and multi-planet systems (which should not have undergone violent migratory processes) around both hot and cool stars will help to discriminate between these hypotheses.

The vast majority of the measurements of the spin-orbit misalignments of transiting exoplanets have come via radial velocity observations of the Rossiter-McLaughlin effect, where distortions in the stellar line profile during the transit are interpreted as an anomalous radial velocity shift. An alternative method, which we utilize, is Doppler tomography, which has been used to probe spin-orbit misalignments for both planets \citep[e.g.,][]{CC189733,CollierCameron10,Brown12,AlbrechtMultis} and stars \citep[e.g.,][]{Albrecht07}. Here, the spectral line profile distortions are spectroscopically resolved and tracked over the course of the transit. The motion of the line profile perturbation during the transit is a probe of the spin-orbit misalignment $\lambda$. While for the most rapidly rotating planet-host stars $\lambda$ can be measured purely from photometry due to the effects of gravity darkening on the surface brightness profile of the star \citep[e.g.,][]{Barnes09,Barnes11}, this method results in a four-fold degeneracy between $\lambda=\pm x^{\circ}$ and $\lambda=180^{\circ}\pm x^{\circ}$. Doppler tomography can break this degeneracy.

In addition to measurements of $\lambda$, Doppler tomography can be used to validate transiting planet candidates around rapidly rotating stars. These stars are not amenable to follow-up using high precision radial velocity observations due to their significantly rotationally broadened stellar lines. Detection of the Doppler tomographic transit signature allows us to verify that the transiting object is indeed orbiting the expected star, i.e., that the system is not a background eclipsing binary blended with a brighter foreground star. By examining the line shape we can also rule out scenarios where the transiting object is another star, as we will be able to see an additional set of absorption lines superposed upon those of the primary. The limitation, however, is that Doppler tomography cannot measure the mass of the transiting object, and thus we cannot distinguish between a hot Jupiter, a brown dwarf, and a small M dwarf. All of these have similar radii and the latter of these would, in many cases, have an insufficient flux ratio to make a detectable imprint upon the visible light spectrum of the primary.

To date the only transiting planet candidate validated using Doppler tomography is WASP-33\,b \citep{CollierCameron10}. There are, however, a number of planet candidates discovered by the {\it Kepler} mission around rapidly rotating stars which can be validated using Doppler tomography. We have begun a program using the telescopes at McDonald Observatory to validate suitable candidates, with a particular focus on longer-period candidates. These will provide a test of the hypotheses described above, as these planets should not have undergone significant tidal damping and so should retain their primordial orbital alignments.

In this paper we describe our Doppler tomography code and present our observations of the hot Jupiter Kepler-13 Ab. Although Kepler-13 Ab has been validated as a planet using Doppler beaming and ellipsoidal variations \citep[e.g.,][]{Shporer11}, it is one of the most favorable {\it Kepler} targets for Doppler tomography and thus presents a good test of our code. 

\section{The Kepler-13 System}

The Kepler-13 (aka KOI-13, BD+46 2629) system has long been known to be a proper motion binary \citep{Aitken04}. \cite{Szabo11} determined that it consists of two A-type stars with similar properties (see Table~\ref{starknowledge}), which are separated by 1.12'' \citep{Adams12}. \cite{Szabo11} also determined that the transiting planet Kepler-13 Ab \citep[detected by ][]{Borucki11} orbits the brighter of the two binary components, Kepler-13 A. Despite the resulting blend, as the separation between Kepler-13 A and B is much smaller than the size of one of \emph{Kepler}'s pixels, the inferred radius for Kepler-13 Ab remains in the planetary range, albeit at the highly inflated end of that range. This is unsurprising, considering the luminous host star and close orbital proximity of the planet to the star, and consequently high planetary temperature. 

\cite{Santerne12} detected a third stellar component in the system in an eccentric binary orbit about Kepler-13 B via the reflex motion of star B. They determined that this companion, Kepler-13 C \citep[denoted Kepler-13 BB by][]{Shporer14}, has a mass of $0.4 M_{\odot}<M<1 M_{\odot}$ and an orbital period of 65.8 days. Kepler-13 Ab thus orbits one member of a stellar triple system; alternatively, due to the massive nature of the planet Kepler-13 Ab, the system could be considered to be a hierarchical quadruple. 

Kepler-13 A is distinguished as one of the hottest stars to host a confirmed planet ($T_{\mathrm{eff}}=8500 \pm 400$ K). 
Stellar parameters for the three stars in the Kepler-13 system are given in Table \ref{starknowledge}, while planetary and transit parameters are summarized in Table \ref{oldknowledge}. As Kepler-13 Ab is a hot Jupiter, it is one of the hottest known planets; \cite{Mazeh12} estimated $T_{\mathrm{eff}}=2600 \pm 150$ K using the secondary eclipse depth in the {\it Kepler} passband.

Kepler-13 Ab was first validated by \cite{Barnes11} through detection of a gravity-darkening signature in the transit lightcurve from \emph{Kepler}. This also enabled them to measure the spin-orbit misalignment, albeit with degeneracies, to be $\lambda=\pm23^{\circ}\pm4^{\circ}$ or $\lambda=\pm157^{\circ}\pm4^{\circ}$. \cite{Shporer11}, \cite{Mazeh12}, \cite{MislisHodgkin12}, \cite{Esteves13}, and \cite{Placek13} detected Doppler beaming and ellipsoidal variations due to the planetary orbit, and used these to measure the mass of Kepler-13 Ab to be ${\sim8-10 M_J}$, putting it firmly below the deuterium burning limit. Many of these different authors, however, found conflicting values for some of the transit and system parameters, especially the impact parameter $b$, ranging from 0.25 to 0.75 (see Table \ref{oldknowledge} for the planetary parameters). While the orbital plane of Kepler-13 Ab has been shown to be precessing, resulting in changes in the transit duration and impact parameter \citep{Szabo12,Szabo14}, the rate of change of the impact parameter found by \cite{Szabo12}, ${db/dt=-0.016 \pm 0.004}$~yr$^{-1}$, is much too small to account for these discrepancies. While \cite{Szabo11} found no evidence for orbital eccentricity, recently \cite{Shporer14} measured a secondary eclipse time offset by $\sim30$ seconds from that expected assuming a circular orbit. This could be caused by either a very small eccentricity ($e\sim5\times10^{-4}$), or a bright spot on the planetary surface offset to the west of the substellar point.

Kepler-13 A is rapidly rotating \citep[$v\sin i=76.6$ km~s$^{-1}$;][]{Santerne12} and bright for a {\it Kepler} target ($Kp=9.96$), making it an excellent target for Doppler tomography. While there is a previous measurement of $\lambda$ via gravity darkening \citep{Barnes11}, as noted above this method cannot distinguish between prograde and retrograde orbits. We can break this degeneracy with Doppler tomography. With this work Kepler-13 Ab becomes the first planet with measurements of $\lambda$ from both photometric and spectroscopic techniques, an important consistency check. Additionally, \cite{Albrecht12} showed that, in addition to the stellar $T_{\mathrm{eff}}$, the planetary scaled semi-major axis $a/R_*$ and mass ratio $M_p/M_*$ are correlated with the degree of alignment. A measurement of the spin-orbit misalignment for Kepler-13 Ab helps to expand the parameter space, as it is a particularly massive planet orbiting close to a massive star.

\begin{deluxetable}{lcc}
\tabletypesize{\scriptsize}
\tablecolumns{3}
\tablewidth{0pt}
\tablecaption{Parameters of Kepler-13 A, B, and C from the Literature \label{starknowledge}}
\tablehead{
\colhead{Parameter} & \colhead{\cite{Santerne12}} & \colhead{\cite{Szabo11}}
}

\startdata
 & System Parameters & \\
$d$ (pc) & \ldots & $500$ \\
$age$ (Gyr) & \ldots & $0.708^{+0.183}_{-0.146}$ \\
$A_V$ (mag) & \ldots & $0.34$ \\
\hline
 & Kepler-13 A  & \\
$V$ (mag) & \ldots & 9.9 \\
$T_{\mathrm{eff}}$ (K) & \ldots & $8511^{+401}_{-383}$ \\
$\log g$ (cgs) & \ldots & $3.9 \pm 0.1$ \\
$[$Fe/H$]$ & \ldots & 0.2 \\
$v\sin i$ (km s$^{-1}$) & $76.6 \pm 0.2$ & $65 \pm 10$ \\
$M_* (M_{\odot})$ & \ldots & 2.05 \\
$R_* (R_{\odot})$ & \ldots & 2.55 \\
\hline
 & Kepler-13 B  & \\
$V$ (mag) & \ldots & 10.2 \\
$T_{\mathrm{eff}}$ (K) & \ldots & $8222^{+388}_{-370}$ \\
$\log g$ (cgs) & \ldots & $4.0 \pm 0.1$ \\
$[$Fe/H$]$ & \ldots & 0.2 \\
$v\sin i$ (km s$^{-1}$) & $62.7 \pm 0.2$ & $70 \pm 10$ \\
$M_* (M_{\odot})$ & \ldots & 1.95 \\
$R_* (R_{\odot})$ & \ldots & 2.38 \\
\hline
 & Kepler-13 C & \\
$P$ (days) & $65.831 \pm 0.029$ & \ldots \\
$e$  & $0.52 \pm 0.02$ & \ldots \\
$K$ (km s$^{-1}$) & $12.42 \pm 0.42$ & \ldots \\
$M_* (M_{\odot})$ & $>0.4, <1$ & \ldots \\
\enddata

\tablecomments{$K$ is the radial velocity semi-amplitude of Kepler-13 B due to its mutual orbit about Kepler-13 C.}

\end{deluxetable}

\section{Methodology}
\label{methodology}

\subsection{Observations}
\label{observationsec}

Observations of Kepler-13 Ab were taken with two telescopes located at McDonald Observatory, the 9.2m Hobby-Eberly Telescope (HET) and the 2.7m Harlan J.\ Smith Telescope (HJST). The HET utilizes a fiber-fed cross-dispersed echelle spectrograph, the High-Resolution Spectrograph \citep[HRS;][]{HRS}. The fibers have a diameter of 2'', and so our observations include blended light from both Kepler-13 A and B \citep[the mutual separation is $1.12$'';][]{Adams12}. This complication is discussed in more detail later in the text. The Robert G. Tull Spectrograph \citep[TS23;][]{Tull95} on the HJST, on the other hand, is a more traditional slit coud\'e spectrograph. There is no facility to correct for image rotation, and so the relative contributions to the spectrum from Kepler-13 A and B vary throughout the course of an observation. While this can, in principle, be corrected for, guiding errors will also cause similar but unpredictable variations. We therefore do not attempt such a correction. Our HRS observations were taken with a resolving power $R=30,000$, while the TS23 observations have $R=60,000$. The spectral range of HRS is $\sim4770$ \AA~to $\sim6840$ \AA, while that of TS23 is $\sim3750$~\AA~to $\sim10200$ \AA; however, none of the orders redward of $\sim8500$~\AA~were  used due to telluric contamination and lack of stellar lines. The exposure time was 300 seconds for all HET observations and 900 seconds for all HJST observations. The mean per pixel signal-to-noise ratio of the continuum is 159 for the HET data and 51 for the HJST data; the mean SNRs for the individual datasets are listed in Table~\ref{observations}.

We observed parts of nine transits of Kepler-13 Ab, seven with the HET and two with the HJST; see Table~\ref{observations}. The transit of 2011 November 5 UT was simultaneously observed with both the HET and the HJST. An additional out-of-transit spectral line template observation was obtained with the HET on 2013 June 28 UT, in order to better determine the out-of-transit line profile.

\begin{deluxetable*}{lcccc}
\tablecolumns{5}
\tablewidth{0pt}
\tablecaption{Observations of Kepler-13 A\lowercase{b} \label{observations}}
\tablehead{
\colhead{Date (UT)} & \colhead{Instrument} & \colhead{Transit Phases Observed} & \colhead{Mean SNR} & \colhead{$N_{\mathrm{spec}}$}
}

\startdata
2011 Jun 8 & HET/HRS & $0.65-0.98$ &   150 &   11\\
2011 Jun 15 & HJST/TS23 & $-0.12-1.25$ &   53 &   16 \\
2011 Jul 6 & HET/HRS & $0.03-0.48$ &   198 &   16\\
2011 Jul 8 & HET/HRS & $0.10-0.51$ &   183 &   15\\
2011 Aug 21 & HET/HRS & $0.21-0.66$ &   162 &   16\\
2011 Sep 13 & HET/HRS & $0.29-0.71$ &   172 &   15\\
2011 Nov 5 & HET/HRS & $-0.09-0.32$ &   135 &   15\\
2011 Nov 5 & HJST/TS23 & $-0.08-0.85$ &   48 &   11\\
2012 Jun 7 & HET/HRS & $0.10-0.60$ &   138 &   17\\
2013 Jun 28 & HET/HRS & template &   120 &   12
\enddata

\tablecomments{We define transit phases such that ingress$=0$ and egress$=1$. The quoted signal-to-noise ratio (SNR) is the SNR per pixel near 5500 \AA. $N_{\mathrm{spec}}$ is the number of spectra obtained during a transit observation.}

\end{deluxetable*}

We perform data reduction using the same IRAF pipelines utilized by the McDonald Observatory Radial Velocity Planet Search Program for HET/HRS \citep[e.g.,][]{Cochran04} and HJST/TS23 \citep[e.g.,][]{Wittenmyer06}. The extracted spectra are then divided by the blaze-profile function, and any residual curvature is removed by fitting a second-order polynomial using a $\sigma$-clipping routine and normalizing.

\subsection{Line Profile Extraction}

The first step in the analysis of the time series line profiles is to extract these line profiles from our spectra. Essentially, we wish to compute the average line profile for each spectrum. We note that in computing an average line profile across a spectrum we ignore variations in the limb darkening parameter as a function of wavelength. As we are interested in the variations in the line profile as a function of time, rather than the detailed line shape, this should not have a significant effect upon our results.

The extraction of the average line profiles from the spectra proceeds in several steps. All steps involve fitting a model spectrum to the data. In all cases this model is produced using the least squares deconvolution method of \cite{Donati97}. In this method, a model spectrum is produced by convolving a model line profile with a series of appropriately weighted delta functions at the wavelengths of the spectral lines. We fit this model spectrum to the data using the least squares methods of \cite{Markwardt09}, as implemented in the IDL function \textsc{mpfit} and derivatives. 

We first select several orders of the spectrum with many telluric lines and few or no stellar lines. We produce a model telluric spectrum using least squares deconvolution using a telluric line list (obtained from the GEISA database\footnote{http://ether.ipsl.jussieu.fr/etherTypo/?id=950}), and assuming a Gaussian line profile. This model spectrum is fit to the data, leaving only the velocity offset between the extracted spectrum and the telluric rest frame as a free parameter. We assume that the telluric rest frame is identical to the spectrograph rest frame ($\pm$ the wind speed, which is much smaller than the velocity scales of interest to us), and so we shift the spectra into this frame. Telluric lines have been shown to be a stable velocity standard \citep[e.g.,][]{GrayBrown06,Figueira10}. The individual spectra display a RMS scatter in the telluric velocities of $\sim250$ m s$^{-1}$, again much smaller than both the velocity scales of interest and the instrumental resolution, although there is a zero-point offset of $\sim6$ km s$^{-1}$ between the spectrograph's intrinsic wavelength calibration and the telluric velocity frame. Now that we have a velocity frame fixed to the Earth, we correct for the Earth's orbital and rotational motion and shift the spectra into the solar barycentric rest frame.

Next we co-add each set of spectra taken on each night, creating several nightly master spectra. For each nightly master spectrum we create a model stellar spectrum. This is produced by obtaining a line list from Vienna Atomic Line Database \citep[VALD;][]{Kupka00}. The line list includes the wavelength of each line, as well as a line depth calculated by VALD using stellar model atmosphere parameters appropriate to our target. We produce an analytic rotationally broadened line profile using Eqn.~18.14 of \cite{Gray}. This profile includes only the effects of rotation; at this stage in the process, we only require an approximately correct line shape. We then fit the model spectrum to each nightly master spectrum, leaving only the velocity offset between the stellar and solar barycentric frames as a free parameter. Now that we have obtained these nightly velocity offsets, we shift all of the spectra into the stellar barycentric rest frame. We note that this assumes that there is no significant acceleration of the star over the course of one night's observations (typically one to a few hours).

As Kepler-13 is a small separation visual binary where one component is itself a single-lined spectroscopic binary, we undertook a small modification to this step for this system. Due to the motion of Kepler-13 B in velocity space, fitting a single line profile results in a bias in the velocity offsets of the spectra that is correlated with the orbital phase of Kepler-13 B. In order to correct for this, we instead fit a model spectrum produced using two analytic rotationally broadend line profiles, with a time-dependent velocity separation given by the orbital elements of \cite{Santerne12}. We determined the contrast between the two profiles by fitting two model line profiles to final extracted line profiles using the unmodified code.

Now that all of the spectra are fixed to the same velocity frame, we co-add all of the out-of-transit spectra to create a template spectrum. We create a model spectrum using the same methodology as described above. Here, however, we fix the velocity offset between model and data at zero and leave the depth of each line as a free parameter. We thus obtain best-fit line depths from our high signal-to-noise template spectrum.

\begin{figure}
\plotone{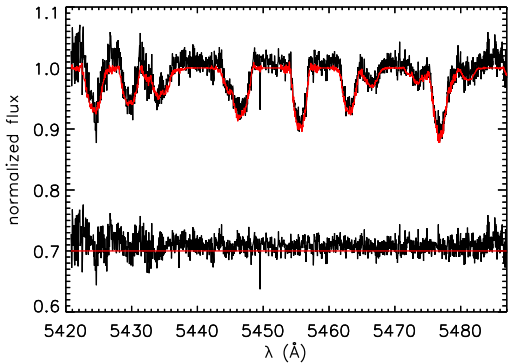}
\caption{One order from one HET spectrum of Kepler-13, showing the final model fit (red in the online journal) to the spectrum (black). The residuals have been shifted upward by 0.7 in order to better show the spectrum. \label{spectrum}}
\end{figure}

The final step is to extract the time series line profiles themselves. For each spectrum we again produce a model spectrum. The line depths are fixed at the best-fit values found earlier. Here the free parameters are the depth of the line profile in each pixel. An example of one of these fits is shown in Fig.\ \ref{spectrum}. For each spectrum we compute the average line profile by computing the weighted mean of the line profiles extracted from each order. Each order's line profile is weighted by the product of the signal-to-noise at the center of that order and the total equivalent width of all lines in that order, after \cite{AlbrechtMultis}. Any orders with noisy line profiles (i.e., the scatter in the continuum is greater than an empirically determined value) are excluded from the computation of the weighted mean. The line profiles from the different orders are also regridded to a common velocity scale. We then perform the same process on the template spectrum to obtain an out-of-transit template line profile. We subtract this template line profile from each of the time series line profiles, resulting in the time series line profile residuals, which display the transit signature.

For Kepler-13, we must again modify this step due to the complicated nature of the system. As will be discussed later, simply subtracting the line profile from our out-of-transit template results in significant systematics because the overall line profile varies as a function of time due to the orbit of Kepler-13 B. In order to correct for this we subtract from each line profile the average line profile from that night of observations. While this subtracts off some of the transit signal, it eliminates almost all of the systematics in the time series line profile residuals.

\subsection{Transit Parameter Extraction}
\label{parextract}

Now that the time series line profile residuals have been computed, we must extract the transit parameters from these data. We compute a model for the time series line profile residuals and fit this to the data. The model is constructed by numerically integrating over the stellar disk, summing the contributions from each surface element to the overall line profile. We divide the stellar disk into approximately 8,000 surface elements.
We utilize Cartesian coordinates for the integration and subsequent computations. We assume a Gaussian line profile with standard deviation 5 km s$^{-1}$ for each surface element; these are then appropriately Doppler shifted, assuming solid body rotation, and scaled by a quadratic limb darkening law. We also neglect macroturbulence; see \S \ref{results} for further discussion of our assumptions on the lack of differential rotation and macroturbulence.

In order to improve computational efficiency, we do not perform the full integration for each exposure. Instead, we first compute the out-of-transit line profile. Then, we compute the location of the planet at the beginning and the end of each exposure (assuming a circular orbit), and for each surface element compute the fraction of the exposure for which that surface element is obscured by the planet. For each surface element, we diminish the out-of-transit line profile by the line profile contribution from that surface element, multiplied by the fraction of the exposure for which that surface element is covered by the planet. Finally, we convolve each line profile with a model instrumental point spread function. 

The steps outlined above are applicable for computing a model for an arbitrary transiting planet. However, for Kepler-13 Ab we need to take some extra care because of the presence of the binary companion Kepler-13 B and its orbit about Kepler-13 C; we must include Kepler-13 B's moving line profile in our model. We use the orbital elements for Kepler-13 B's orbit about Kepler-13 C presented by \cite{Santerne12} to calculate the velocity of Kepler-13 B at each exposure. We then compute a rotationally broadened line profile for Kepler-13 B using the model described above, Doppler shift it and scale it relative to the Kepler-13 A profile, and add it to the line profile for Kepler-13 A. Including this profile and the resulting dilution of the spectroscopic transit signature is necessary to accurately model the data.

Ideally, we would simply fit for all relevant parameters ($\lambda$, $b$, $v\sin i$) simultaneously. As our time series line profiles are derived from the average of many lines across a wide region of the spectrum, and the limb darkening and therefore the detailed line shape change as a function of wavelength, our model line profiles do not fit the average line profile to better than a few percent in the wings of the profile. This poses difficulties for extracting $v\sin i$, as well as the transit parameters. We therefore adopted a two-stage fitting process, first extracting $v\sin i$ from a single line and then $\lambda$ and $b$ from the time series line profile residuals.

For each sequential parameter extraction we used a Markov chain Monte Carlo (MCMC) to sample the likelihood function of the model fits to the data. In all cases we used four chains each of 150,000 steps, cutting off the first 20,000 steps of burn-in. 
In addition to our free parameters for each fit, we also wished to incorporate prior knowledge from the literature, e.g.\ on the transit duration for Kepler-13 Ab. We thus set Gaussian priors upon these parameters; that is, assuming that the errors are Gaussian, we can define an ``effective'' $\chi^2$ statistic
\begin{equation}
\chi^2_{eff}=\sum_{i}\frac{(O_i-C_i)^2}{\sigma_i^2}+\sum_j\frac{(P_j-P_{j,0})^2}{\varsigma_j^2}
\end{equation}
where $O$ denotes the data, $C$ the model, $\sigma$ the calculated error on each data point, $P_j$ the value of parameter $j$ at the given iteration of the Markov chain, $P_{j,0}$ the value of parameter $j$ from the literature, and $\varsigma_j$ the uncertainty on parameter $j$ from the literature, and we are summing over $i$ data points and $j$ model parameters where we have prior information.

First, we model a single line, the Ba \textsc{ii} line at $\lambda$6141.7~\AA, chosen because it is deep but unsaturated and isolated. We fit the nightly master spectra with models of the line profiles of Kepler-13 A and B, neglecting any contribution from the transiting planet. We leave the $v\sin i$ of each star, the contrast between the two stars, and eight nightly velocity offsets as free parameters. We set Gaussian priors upon two quadratic limb darkening parameters for each star, each with a width 0.1, and upon the five parameters determining the radial velocity variation of Kepler-13 B ($P$, epoch, $e$, $\omega$, $K$). For the limb darkening coefficients we use coefficients in the Sloan {\it r} band (as this is the closest standard photometric band to the Ba \textsc{ii} $\lambda$6141.7~\AA\ line), taken from the tables of \cite{Claret04} for an ATLAS model atmosphere and interpolated to the stellar parameters of Kepler-13 A and B as presented by \cite{Szabo11} using the JKTLD code\footnote{http://www.astro.keele.ac.uk/jkt/codes/jktld.html}. We use the methods of \cite{Kipping13} to obtain even sampling in limb darkening space. For the orbital parameters, we set the initial value and prior width to the best-fit value and 1-$\sigma$ uncertainty, respectively presented by \cite{Santerne12}; see Table \ref{starknowledge}.

Second, we fit the time series line profile residuals with an appropriate model using another MCMC. Here we leave $\lambda$ and $b$ as free parameters, and set priors on the $v\sin i$ of Kepler-13 A and contrast between Kepler-13 A and B (with the prior value and width set to the median values and 1-$\sigma$ uncertainty, respectively, on these parameters from the first MCMC), and the limb darkening coefficients of Kepler-13 A, transit depth $R_p/R_*$, transit duration, planetary orbital period, and planetary orbital epoch, with all values and uncertainties/prior widths taken from \cite{Esteves13}. We fix the $v\sin i$ and orbital parameters of Kepler-13 B at values from our first MCMC and \cite{Santerne12}, respectively, in the interests of computational efficiency and as uncertainties in these parameters should have a minimal effect on the line profile residuals.

We note that in principle it is possible to measure the time of mid-transit and the transit duration directly from the spectroscopic data. Additionally, $R_p/R_*$ and $(R_p/R_*)^2$ may be measured independently (the width of the transit signature depends on $R_p/R_*$, while the area under the transit signature is proportional to $(R_p/R_*)^2$). If a system is affected by dilution, the measured value of $(R_p/R_*)^2$ will be smaller than that inferred from the measurement of $R_p/R_*$ from the transit signature width, which is unaffected by dilution. In practice, however, given finite spectral resolution, limited time resolution, and relatively low signal-to-noise, these values are best determined from {\it Kepler} photometry. We thus incorporate these parameters via priors in our MCMCs.

In our second set of MCMCs, we fit the model directly to the time series line profile residuals. Alternatively, we also use a method of binning the spectra to increase the signal-to-noise ratio. This method rests upon the following observation. Neglecting differential rotation of the star and assuming a circular orbit for the planet, the rate of motion of the planetary transit signature across the line profile ($dv/dt$) will be constant. Given the transit duration, each value of $dv/dt$ corresponds to a single value of the velocity difference between the locations of the transit signature at ingress and egress, $v_{14}$. In geometrical terms, the path of the planetary transit signature in the time series line profile residual plots will be a straight line. For a given value of $v_{14}$, the transit signature will occur at some velocity $v_i$ in the $i^{th}$ spectrum. We shift each of the $i$ line profile residuals by $-v_i$, such that the transit signature will occur at the same velocity for each shifted line profile residual, and then bin together all of the shifted line profile residuals. If we have the correct value of $v_{14}$, the transit signatures in each line profile residual will tend to add constructively, and we will obtain a single high signal-to-noise transit signature. If we have an incorrect value of $v_{14}$, the transit signatures will not add coherently, and the diluted transit signature will be below the noise floor. We define the velocity scale of the shifted line profile residuals such that it is $v_{\mathrm{cen}}$, the velocity of the transit signature at the transit midpoint. 

For a grid of possible values of $v_{14}$ ($|v_{14}|\leq 2v\sin i$), we perform this shifting and binning operation, and visualize this as a two-dimensional map of the deviation from the out-of-transit line profile as a function of $v_{\mathrm{cen}}$, $v_{14}$. We model these shifted and binned data by producing model time series line profile residuals in the same manner as above, and then shifting and binning these in the same manner as we have treated the data. We then extract transit parameters from the shifted and binned data using an MCMC similar to the one for the unbinned data described above. While mathematically a complicated, usually double-valued relationship exists between ($\lambda$, $b$) and ($v_{\mathrm{cen}}$, $v_{14}$), qualitatively there exists a simple relationship between ($v_{\mathrm{cen}}$, $v_{14}$) and the path of the transit signature across the stellar disk. For solid-body rotation, and defining a coordinate $x$ on the visible disk of the star perpendicular to the projected stellar rotation axis, each velocity on the line profile maps to a single value of $x$, i.e., $v\propto x$ \citep{Gray}. $v_{\mathrm{cen}}$ and $v_{14}$ together fix the $x$ coordinates of ingress and egress, $x_1$ and $x_4$, respectively. For each pair of $x_1$, $x_4$ there are two possible paths across the stellar disk: one with low $\lambda$, high $b$ and one with high $\lambda$, low $b$, resulting in the double-valued function that maps ($\lambda$, $b$) to ($v_{\mathrm{cen}}$, $v_{14}$). 
 In general, positive values of $v_{14}$ correspond to $|\lambda|<90^{\circ}$, and $v_{14}<0$ corresponds to $|\lambda|>90^{\circ}$, while $v_{\mathrm{cen}}>0$ corresponds to $\lambda>0^{\circ}$ and $v_{\mathrm{cen}}>0$ corresponds to $\lambda<0^{\circ}$.

\begin{figure}
\plotone{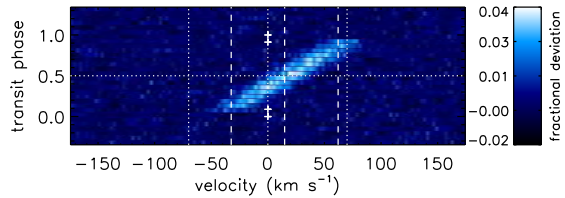}
\caption{Model time series line profile residuals, illustrating $v_{14}$ and $v_{\mathrm{cen}}$. The transit signature is the bright streak moving from lower center to upper right. The three vertical dashed lines mark, from left to right, $v_1$, $v_{\mathrm{cen}}$, and $v_4$, the velocity of the transit signature at ingress, mid-transit, and egress, respectively; $v_{14}=v_4-v_1$. Time increases from bottom to top. The transit phase is defined such that ingress=0 and egress=1. Vertical dotted lines mark $v=0, \pm v\sin i$, and a horizontal dotted line marks the time of mid-transit. Small crosses mark the times of first, second, third and fourth contacts. The units of the color scale are fractional deviation from the average out-of-transit line profile. Note that, in general ($b\neq0$), $v_{\mathrm{cen}}\neq0$. The model was computed for a planet with $\lambda=45^{\circ}$ and $b=0.3$ orbiting a star with $v\sin i=70$ km s$^{-1}$. A small amount of noise has been added to the model to better approximate an actual observation. \label{diagrammatic}}
\end{figure}

\subsection{Testing the Code: WASP-33 b}
\label{testing}

\begin{figure}
\plotone{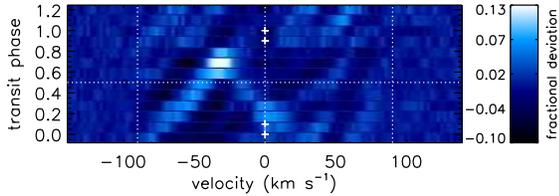}
\caption{Time series line profile residuals of a transit of WASP-33 b; compare to Fig.\ 4 of \cite{CollierCameron10}. Notation on the plot is the same as for Fig. \ref{diagrammatic}. The transit signature is the bright streak moving from bottom center to upper left, while the pattern of alternating dark and light streaks moving from lower left to upper right are non-radial oscillations of the host star WASP-33 \citep[the star is a $\delta$ Sct variable;][]{Herrero11}. \label{wasp33}}
\end{figure}

In order to verify that our code is working correctly, we analyzed one of the Doppler tomographic datasets on WASP-33 b presented by \cite{CollierCameron10}. These observations were taken using the HJST on 2008 November 12 UT. We are able to reproduce their results (Fig.\ \ref{wasp33}), an important test of our code. We measure the quality of the data by the root-mean-squared (RMS) scatter of the continuum; for our WASP-33 data, this amounts to 0.010 of the depth of the line profile. \cite{CollierCameron10} did not provide a quantitative measure of the noise level in their data, but qualitatively our noise floor appears to be somewhat lower than that of the previous work.

We furthermore find a best-fitting model using our MCMC. For WASP-33 there are variations of the line shape of a few percent due to non-radial pulsations of the host star, and so, unlike for Kepler-13 A, we are able to model the line shape to within the uncertainties from the pulsations. We thus conduct only a single MCMC, fitting for $v\sin i$, $\lambda$, and $b$ simultaneously. We use limb darkening coefficients interpolated to the stellar parameters from \cite{CollierCameron10} using JKTLD, but here use the \cite{Claret00} values for the $V$ band. We obtain values of $v\sin i=87.4 \pm 0.2$ km s$^{-1}$, ${\lambda=-111.2^{\circ} \pm 0.3^{\circ}}$, and $b=0.1738 \pm 0.0043$. Note that these uncertainties take into account only statistical errors and do not include systematic errors, which will be discussed later for the case of Kepler-13 Ab. 
Working from the McDonald data, \cite{CollierCameron10} obtained $v\sin i=85.64\pm0.13$  km s$^{-1}$, $\lambda=-105.8^{\circ}\pm1.2^{\circ}$, and $b=0.176\pm0.010$.  We attribute the differences between our measured parameters and those of \cite{CollierCameron10} to the complication of the stellar non-radial pulsations.

\begin{figure}
\plotone{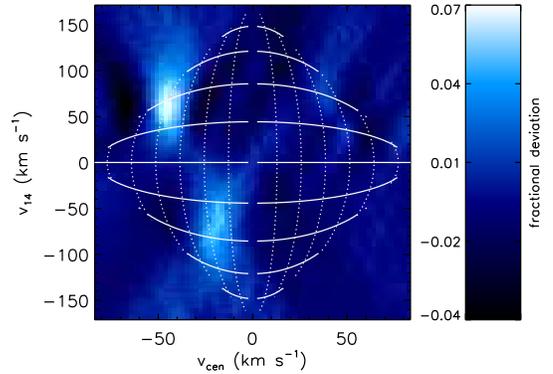}
\caption{Time series line profile residuals of a transit of WASP-33 b, shifted and binned according to the scheme described in the text. $v_{\mathrm{cen}}$ is the velocity of the transit signature at the transit midpoint, while $v_{14}$ is the difference between the velocity of the transit signature at egress and ingress. Two bright peaks are apparent; the one at bottom center is the transit signature, while the one at upper left is due to the most prominent of the non-radial oscillations. Other structures in the map are also due to the non-radial oscillations. The solid lines show lines of constant $\lambda$, while the dotted show lines of constant $b$. The $\lambda$ contours mark, from top to bottom, $\lambda=\pm30^{\circ}, \pm45^{\circ}, \pm60^{\circ}, \pm75^{\circ}, \pm90^{\circ}, \pm105^{\circ}, \pm120^{\circ}, \pm135^{\circ}, \pm150^{\circ}$ ($\lambda$ is positive on the right half of the plot, and negative on the left half). The $b$ contours mark, from the centerline of the plot outwards, $b=0.15, 0.30, 0.45, 0.60, 0.75, 0.9$. Note that the transit signature lies between the $\lambda=-105^{\circ},-120^{\circ}$ and the $b=0.15, b=0.30$ contours, as we would expect. We note that the relationship between ($v_{\mathrm{cen}}, v{_14}$) and ($\lambda, b$) is double-valued; only the solution appropriate to WASP-33 b is shown here. \label{wasp33shift}}
\end{figure}

We also shift and bin our WASP-33 data, as described above. The resulting map is shown in Fig.\ \ref{wasp33shift}. There are two strong peaks in the map, one due to the planetary transit and the other due to non-radial pulsations. We attempted to extract transit parameters from these data using our MCMC, but due to the non-radial pulsations we could not obtain a satisfactory fit.

\section{Results}
\label{results}

For Kepler-13, using our first MCMC we measure projected rotational velocities for the two stars of ${v\sin i_A=76.96 \pm 0.61}$ km s$^{-1}$ and $v\sin i_B=63.21 \pm 1.00$ km s$^{-1}$, which agree to within $1\sigma$ with the $v\sin i$ values presented by \cite{Santerne12}.

\begin{figure}
\plotone{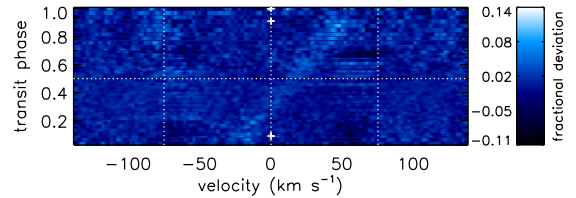}
\caption{Transit signature of Kepler-13 Ab, using the best quality HET data (all transits except those of 2011 Nov 5 and 2012 Jun 7, which were excluded due to lower signal-to-noise; see Table~\ref{observations}). The transit signature is the bright streak moving from lower left to upper right. Note the large ($\sim 0.1$ of the depth of the line profile) systematics. Notation on the figure is the same as on Fig.\ \ref{wasp33}. \label{koi13_avg}}
\end{figure}

\begin{figure}
\plotone{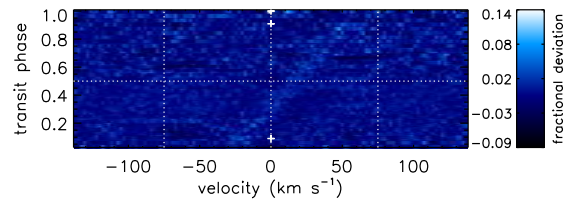}
\caption{Same as Fig.\ \ref{koi13_avg}, except subtracting off the average line profile from each night. Note that most of the systematics have vanished, but the amplitude of the transit signature has also been reduced. Notation on the figure is the same as on Fig.\ \ref{wasp33}. \label{koi13_nights}}
\end{figure}

\begin{figure}
\plotone{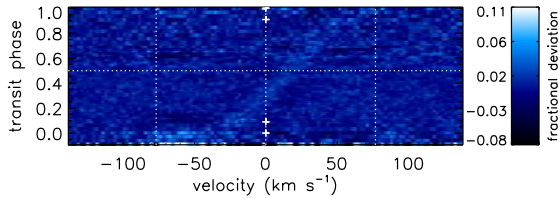}
\caption{Transit signature of Kepler-13 Ab, subtracting off the average line profile from each night and using all of our data. For display purposes points with fractional deviations from the out of transit line profile greater than 0.11 or less than -0.08 have been set to these values, in order to better display the transit signature. This only affects the earliest spectrum. Notation on the figure is the same as on Fig.\ \ref{wasp33}. \label{koi13joined}}
\end{figure}

\begin{figure}
\plotone{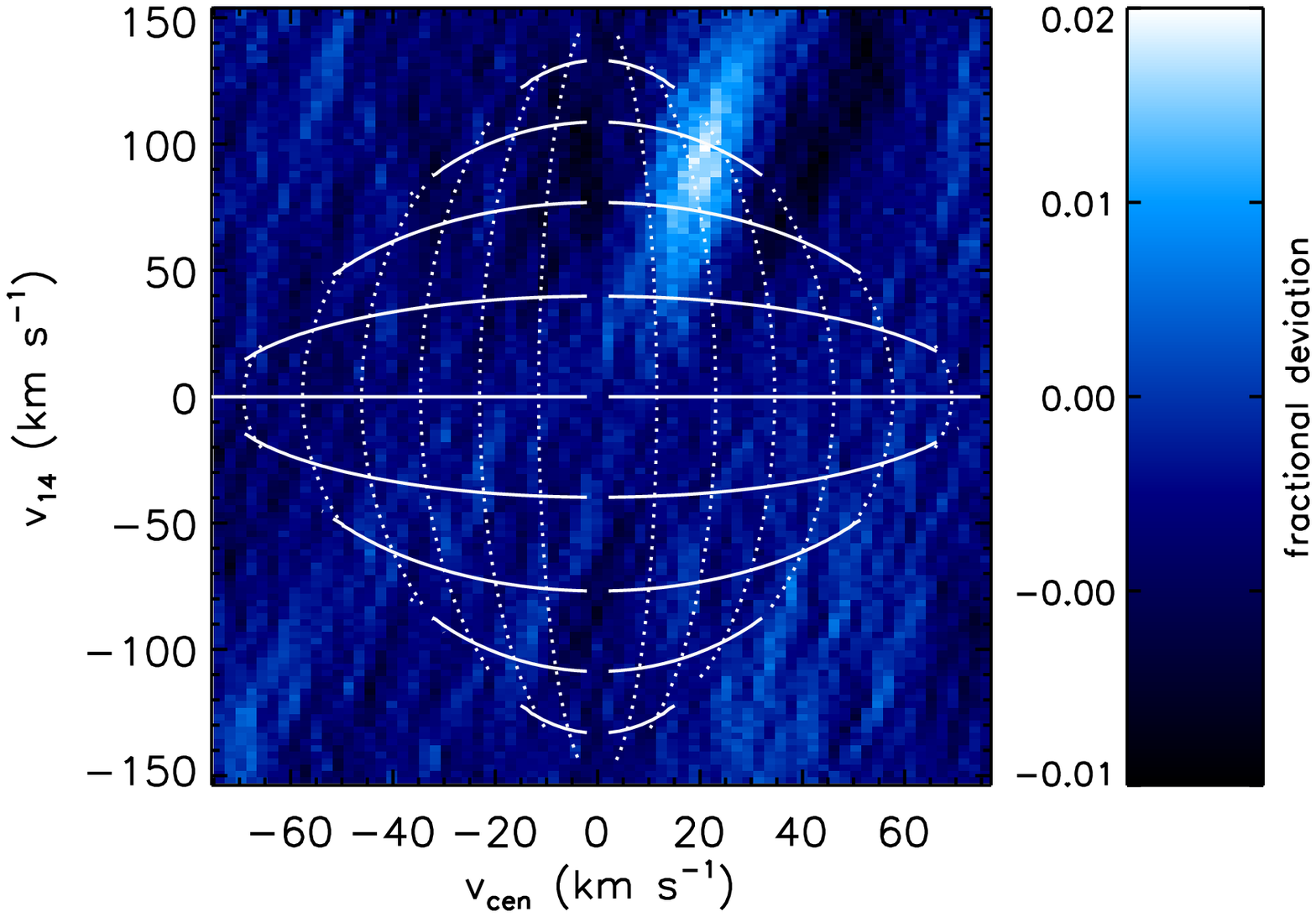}
\plotone{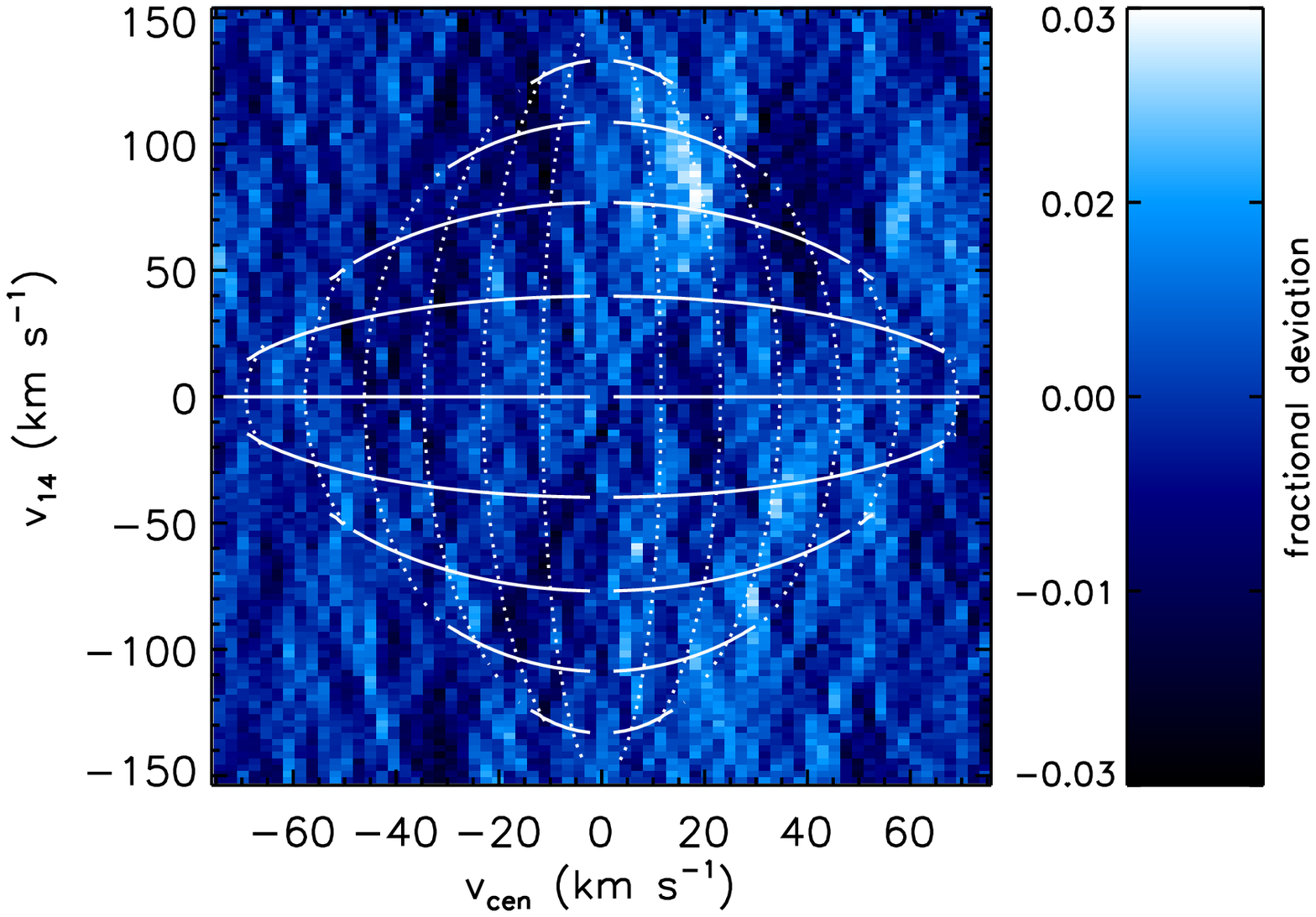}
\caption{Top. Transit data from the HET, binned according to the scheme discussed in the text. Bottom. Same as top, but for the HJST data. A bright spot is visible in the same location as in the HET data, indicating a low signal-to-noise detection of the transit. The contours are the same as in Fig.~\ref{wasp33shift}. The dark sidelobes on either side of the bright transit signature (especially prominent in the HET data, top) are the result of subtracting off the average line profile from each night, rather than an out-of-transit line profile. \label{koi13binned}}
\end{figure}

In Fig.~\ref{koi13_avg} we show the time series line profiles extracted from the HET data, produced by subtracting the out-of-transit template line profile from each of the time series line profiles. Significant systematics are visible, of amplitude $\sim0.1$ of the depth of the line profile. Most of these systematics result from differences between the time series line profiles and the out-of-transit template line profile due to the motion of Kepler-13 B in velocity space. This is illustrated in Fig.~\ref{koi13_nights}, where we have subtracted the average line profile {\it from each night} from each of the time series line profiles. Fig.~\ref{koi13joined} is identical to Fig.~\ref{koi13_nights}, except using all of our HET data. Due to these systematics, for the remainder of the analysis we subtract the nightly average line profile from the time series line profiles, and we do not use the out-of-transit template data. The RMS scatter of the continuum is 0.022 times the line depth. The transit signature is immediately apparent visually. That the planetary orbit is prograde can be determined by inspection, as the transit signature is over the blueshifted hemisphere of the star at ingress and moves across to the redshifted hemisphere by egress. We also shift and bin the HET data (see Fig.\ \ref{koi13binned}, top). Again, the transit signature is clearly detected.

\begin{figure}
\plotone{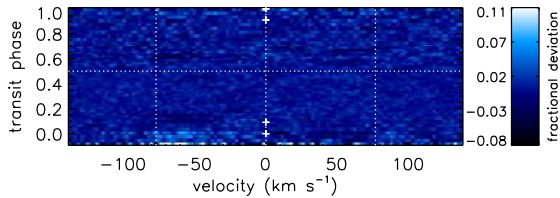}
\caption{Same as Fig.\ \ref{koi13joined}, but with the best-fitting transit model subtracted. The transit signature is well subtracted. For display purposes points with fractional deviations from the out of transit line profile greater than 0.11 or less than -0.08 have been set to these values, in order to better display the transit signature. This only affects the earliest spectrum. Notation on the figure is the same as on Fig.\ \ref{wasp33}. \label{koi13resids}}
\end{figure}

Our best-fit values and 1-$\sigma$ uncertainties from the MCMCs are shown in Table \ref{values}. We present values from both directly fitting the data and fitting the shifted and binned data; these two methods give consistent results. The binned data have smaller uncertainties, but in order to be conservative and as the direct fits have a reduced chi-squared closer to 1 ($\chi^2_{red}=1.13$ for the direct fit, $\chi^2_{red}=0.66$ for the shifted and binned fit),  we quote these values. 
We find a best-fit spin-orbit misalignment of $\lambda=58.6^{\circ} \pm 1.0^{\circ}$, in disagreement with the value of $\lambda=23^{\circ}\pm4^{\circ}$ found by \cite{Barnes11}. We also find $b=0.256 \pm 0.011$. We note that the quoted uncertainties on these parameters are the formal statistical uncertainties, given the assumptions made in our models. They do not include systematic uncertainties, which we discuss in detail later in this section. In Fig.\ \ref{koi13resids} we show the time series line profile residuals with the best-fit model, using these parameters, subtracted off. 

\begin{figure}
\plotone{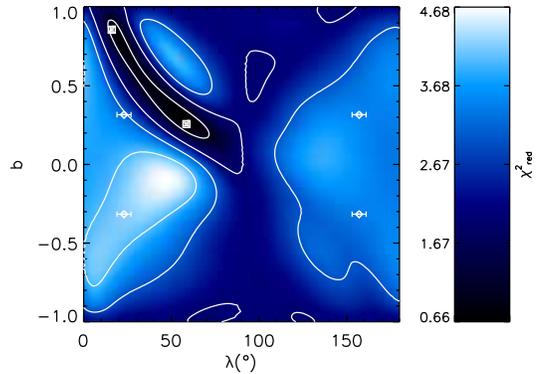}
\caption{Reduced $\chi^2$ space for our shifted and binned data, in $\lambda$ and $b$. The four solutions allowed by \cite{Barnes11} and their associated uncertainties are marked by diamonds; \cite{Barnes11} did not quote an uncertainty on their value of $b$. The two best-fit solutions allowed by our data are denoted by squares. For this display we allow negative values of $b$; note that a transit chord with ($+\lambda, -b$) is identical to one with ($-\lambda, +b$). The contours denote $\chi^2_{\mathrm{red}}=1, 2, 3, 4$. \label{chi2space}}
\end{figure}

We note that our data also permit a second solution, with $\lambda=16.04^{\circ} \pm 0.72^{\circ}$ and $b=0.856 \pm 0.014$. This solution, however, has a slightly worse value of reduced chi-squared ($\chi^2_{red}=1.03$) and moreover implies a physically unrealistically low value for the stellar mean density, $\bar{\rho}_*=0.04$ g cm$^{-3}$. We calculated the stellar mean density using Eqn.~9 of \cite{SeagerMallenOrnelas03}, which is, using the nomenclature used in this article,
\begin{equation}
\bar{\rho}_*=\bigg(\frac{4\pi^2}{P^2G}\bigg)\bigg(\frac{(1+R_p/R_*)^2-b^2[1-\sin^2(\tau_{14}\pi/P)]}{\sin^2(\tau_{14}\pi/P)}\bigg)^{3/2}
\end{equation}
where $P$ is the planetary orbital period and $\tau_{14}$ is the transit duration, both measured from {\it Kepler} photometry. Note that the inferred stellar mean density depends only upon our measurement of $b$ and does not directly depend upon $\lambda$. Given this stellar mean density and the stellar surface gravity measured by \cite{Szabo11} ($\log g=3.9\pm0.1$), we have two independently-measured parameters which physically depend only on the stellar mass and radius; thus, we can estimate the stellar mass and radius implied by $\bar{\rho}_*$ and see whether it is compatible with the other system parameters. A value of $\bar{\rho}_*=0.04$ g cm$^{-3}$ implies a stellar radius of $R_*=8-13 R_{\odot}$ and mass of $M_*=15-60 M_{\odot}$, parameters which are incompatible with the \cite{Szabo11} value of $T_{eff}=8511^{+401}_{-383}$ K, as well as the other measured parameters of the system. Performing the same exercise for $b=0.256$ results in a stellar mass and radius consistent with those found by \cite{Szabo11} and \cite{Barnes11}. The full $\chi^2$ space for our data is shown in Fig.~\ref{chi2space}.

\begin{figure}
\plotone{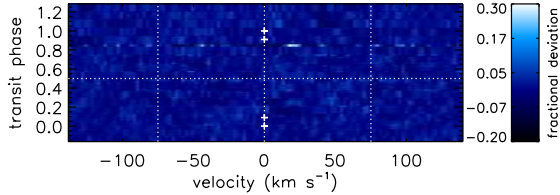}
\caption{Transit data on Kepler-13 Ab from the HJST, using data from both observed transits. The transit signature is not apparent to the eye. Notation on the figure is the same as on Fig.\ \ref{wasp33}. \label{koi13hjst}}
\end{figure}

We also observed two transits of Kepler-13 Ab using the HJST. These data are shown in Fig.\ \ref{koi13hjst}. Like for the HET, in order to produce the time series line profile residuals, we subtract off the average line profile from each night rather than an out-of-transit line profile from both nights. The data are at a much lower signal-to-noise level than our HET data (the RMS scatter of the normalized continuum is 0.037 times the line depth), and the transit is not readily apparent to the eye in the time series line profile residual map. We apply the bin-and-shift method to the HJST data (see Fig.\ \ref{koi13binned}, bottom). Here, we recover the same transit signature seen in the HET data, albeit at lower signal-to-noise. Here we measure values of $\lambda=60.5^{\circ} \pm 1.1^{\circ}$ and $b=0.168 \pm 0.010$. The spin-orbit misalignment is in mild disagreement with the value from the direct fit to the HET data, at a level of $1.3\sigma$ for $\lambda$, while there is a strong $6\sigma$ disagreement between the impact parameter found from the HET and HJST data. One possible cause is the varying degree of contamination from Kepler-13 B during the observations due to field rotation (as noted above, the TS23 is a slit spectrograph). Another possible cause is the poorer time resolution of the HJST data as compared to the HET (exposure times were 900 s for the HJST and 300 s for the HET). In the spectroscopic data the impact parameter is constrained, in part, by how quickly the transit signature increases (decreases) between first and second (third and fourth) contacts. Thus, the lower time resolution of the HJST could introduce larger systematic uncertainties in these data. Additionally, the values above include only statistical uncertainties, which overstate the true degree of discrepancy between the HET and HJST values. We have, however, been unable to positively identify the source of this discrepancy.

The formal uncertainties on our values for $\lambda$ and $b$ quoted earlier are the statistical uncertainties given the assumptions that we have made in our models (no differential rotation or microturbulence, etc.) and do not contain information on systematic sources of uncertainty, which we will now discuss. 

One possible source of systematic errors is the presence of differential rotation, which we have neglected in our models. \cite{AmmlervonEiffReiners12} analyzed the line profiles of A and F dwarfs for evidence of differential rotation. They found no stars with $T_{\mathrm{eff}}\gtrsim8500$ K that exhibited differential rotation. \cite{Balona13}, however, used Fourier analysis of the {\it Kepler} lightcurves of A stars to infer that these stars exhibit a similar degree of differential rotation to the sun. We constructed a modified version of our models that include differential rotation, and conducted a version of our first MCMC, fitting to the line profile shape, in order to constrain the differential rotation. For a differential rotation law $\omega=\omega_0-\omega_1\sin^2\phi$, where $\phi$ is the latitude on the stellar surface, the differential rotation parameter $\alpha$ can be defined as $\alpha=\omega_1/\omega_0$ \citep[for the Sun, $\alpha=0.20$;][]{ReinersSchmitt02}. We note that we also need to include the stellar inclination $i$ with respect to the line of sight in this model; however, we find $i$ to be totally unconstrained. The results of this exercise indicate the presence of a small amount of differential rotation. Overall, we find $\alpha=0.050 \pm 0.028$; however, there does exist a degeneracy such that higher values of $|i|$ result in larger preferred values for $\alpha$: we find $\alpha=0.034 \pm 0.017$ for $i=0^{\circ}$ and $\alpha=0.046 \pm 0.023$ for $|i|=48^{\circ}$, the value found by \cite{Barnes11}. This is consistent with the results of \cite{Szabo14}, who found splitting of the frequency spectrum peak associated with rotation, likely due to differential rotation.

In order to test the effects of this level of differential rotation on our measurement of the transit parameters, we modified our second MCMC to include differential rotation. We added two parameters, $\alpha$ and the stellar inclination $i$. $i$ was allowed to float, while, due to the dependence of the best-fit $\alpha$ on $i$, we included a variable prior on $\alpha$ depending on the value of $i$. Marginalizing over $i$ in 5$^{\circ}$ bin sizes, we found the mean and standard deviation of $\alpha$ for each bin and used these as the prior center and width for the new MCMC. From this MCMC, we obtain $\lambda=56.56^{\circ} \pm 0.85^{\circ}$ and $b=0.2870 \pm 0.0095$. We note that the presence of even strong differential rotation cannot bring our value of $\lambda$ into agreement with that found by \cite{Barnes11}.  

We also neglected macroturbulence in our models, which could potentially induce systematic uncertainties in our measured values of $\lambda$, $b$. Measurements of macroturbulence in A dwarfs in the literature are somewhat lacking. \cite{SimonDiazHerrero13} found varying degrees of macroturbulent broadening for B dwarfs, ranging from none to several tens of km s$^{-1}$ (they note that this ``macroturbulence'' is not necessarily physical turbulence). \cite{Fossati11} measured macroturbulent broadening of order $\sim10$ km s$^{-1}$ for two late A dwarfs. \cite{Aerts09} argued that ``macroturbulence'' in early-type stars is actually due to the collective action of many low-amplitude pulsational modes; early-type stars which do not pulsate should not show this type of macroturbulence. Even with {\it Kepler}'s photometric precision, there is little evidence for any pulsation of Kepler-13 A which could result in this type of macroturbulence. \cite{Cantiello09} conducted simulations of convection in the outer layers of massive stars due to an opacity peak produced by Fe ionization. They found that such zones can cause surface granulation and consequent small-scale velocity fields in stellar photospheres. They find, however, that this effect does not occur for stars with $L<10^{3.2} L_{\odot}$ for Galactic metallicities, and is furthermore more prominent at low surface gravities. As Kepler-13 A is below this luminosity cutoff ($L=10^{1.5} L_{\odot}$) and has high surface gravity \citep[$\log g=3.9 \pm 0.1$;][]{Szabo11}, we conclude that surface granulation due to this mechanism should not occur for Kepler-13 A.

A key question for estimating the effects of macroturbulence upon our results lies with the scales of macroturbulent velocity fields in the stellar atmosphere. If these scales are much smaller than the size of the projected planetary disk during the transit, then this will simply increase the range of radial velocities over which the planet subtracts light from the line profile. The effect will be to ``smear out'' the transit signature, but this should not affect the measured value of $\lambda$. If, however, the macroturbulent velocity field changes on scales of similar or greater size as the planetary disk, then the velocity of the region of the stellar disk covered by the planet will differ from that expected if taking only rotation into account. Thus, the planetary transit signal in each spectrum will exhibit a quasi-random shift from the expected velocity.

\cite{KallingerMatthews10} presented evidence that some of the large number of frequencies seen in the frequency spectra of $\delta$ Sct (early A) stars observed by CoRoT are in fact due to surface granulation rather than pulsations, as pulsations at these frequencies would be of such high degree $l$ that they should not be evident in integrated disk photometry. Based upon the inferred granulation frequencies, they find that the granulation properties follow scaling laws derived for solar-type stars. When scaling from such solar models, \cite{Stello07} make the assumption that the size of granulation cells is proportional to the atmospheric pressure scale height $H_P$. \cite{KjeldsenBedding95} use the scaling relation $H_P\propto T_{eff}/g$. Using these relations and the stellar properties of Kepler-13 A from \cite{Szabo11}, we estimate that the size of any surface granulation cells for Kepler-13 A should be $\sim5$ times that of such cells on the Sun, or $\sim0.1 R_J$, comfortably below the size scale of the planetary disk \citep[using an average solar granule size of 1300 km, from][]{Gray}. Nonetheless, given the uncertainty in the relations used to derive this estimate, we choose to include ``jitter'' caused by large-scale macroturbulent cells in the stellar atmosphere in our MCMCs (note that this is not the same as the jitter frequently invoked as a source of noise in radial velocity observations).

In order to simulate the effect of macroturbulence on the size scale of the planet, we use the following approach. We allow each of the time series line profile residuals to have a small velocity offset from its nominal value. The effect of this is to shift the transit signature in that line profile residual in velocity space. Since we have already subtracted off the average line profile shape, this mimics a velocity shift of the transit signature due to large-scale macroturbulence rather than a radial velocity offset for the entire line profile. For computational reasons we apply this velocity shift to the model line profile residuals, not the data. At each MCMC step, we perform a single parameter minimization for each velocity offset using MPFIT. Similar methodologies have been used by \cite{AlbrechtMultis} to deal with jitter and by \cite{Albrecht14} to handle stellar pulsations. We limit the velocity offset amplitude to 15 km s$^{-1}$ in order to prevent the model transit signatures from latching on to the remaining systematics in the data. The mean offset amplitude is 5.7 km s$^{-1}$. From these MCMCs, we obtain $\lambda=60.4^{\circ} \pm 1.6^{\circ}$ and $b=0.230 \pm 0.030$.

We thus find that including ``jitter'' and differential rotation have opposite systematic effects on our results: large-scale macroturbulence shifts the best-fit parameters to higher $\lambda$ and lower $b$, while differential rotation shifts them to lower $\lambda$ and higher $b$. Thus, we expect that these effects should largely cancel each other out, and our overall result should not be affected, while increasing the uncertainty in our results. In order to remain $1\sigma$ consistent with both the differential rotation and ``jitter'' MCMC results, we therefore adopt $\lambda=58.6^{\circ} \pm 2.0$ and $b=0.256 \pm 0.030$.

Additionally, our model assumes an intrinsic line standard deviation of 5 km s$^{-1}$. In order to test the impact of this assumption on our results we fit a model with an intrinsic line standard deviation of 10 km s$^{-1}$ to our data. This did not significantly alter our measured values of $\lambda$ and $b$ or the $\chi^2_{\mathrm{red}}$ value of the model fits, and so we conclude that this has minimal impact on our measurements.

\begin{deluxetable*}{cccccc}
\tablecolumns{5}
\tablewidth{0pt}
\tablecaption{Best-Fit Values for Kepler-13 A\lowercase{b} Parameters \label{values}}
\tablehead{
\colhead{Parameter} & \colhead{Adopted} &\colhead{HET direct fit} & \colhead{HET binned fit} & \colhead{HJST binned fit} 
}

\startdata
$v\sin i_A$ (km s$^{-1}$) &   \ldots& $76.96 \pm 0.61$ & \ldots  & \ldots \\
$v\sin i_B$ (km s$^{-1}$) &  \ldots& $63.21 \pm 1.00$ & \ldots & \ldots \\
\hline
$\lambda$ ($^{\circ}$) & $58.6 \pm 2.0$ & $58.6 \pm 1.0$ & $58.24 \pm 0.68$ & $60.5 \pm 1.1$ \\
$b$ & $0.256 \pm 0.030$ & $0.256 \pm 0.011$  & $0.266 \pm 0.007$ & $0.168 \pm 0.010$ 

\enddata

\tablecomments{The quoted uncertainties for all except the ``adopted'' column are the formal statistical uncertainties and do not take systematic uncertainties into account.}

\end{deluxetable*}

\begin{figure}
\plotone{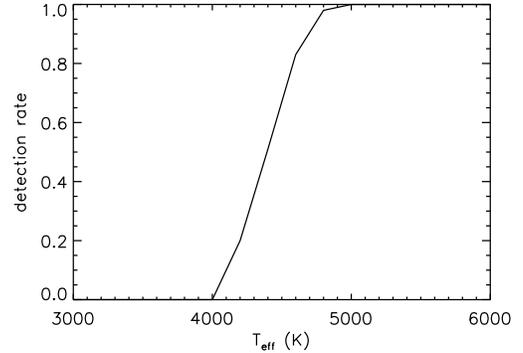}
\caption{Detection rate of synthetic spectral signals of Kepler-13 C injected into our data as a function of the effective temperature of the companion, assuming that it is a main sequence star. \label{Csensitivity}}
\end{figure}

In addition to detecting the transit signal of the planet Kepler-13 Ab, we set upper limits on the mass of the tertiary stellar companion Kepler-13 C. We follow \cite{GulliksonDodsonRobinson} to cross-correlate all HET spectra against model spectra of late-type stellar companions and search for significant cross-correlation function (CCF) peaks. Since the orbit of Kepler-13 B is known \citep{Santerne12}, we can predict the velocity of Kepler-13 C by assuming some guess mass. We can then shift the CCFs by that velocity and co-add them, amplifying any CCF peak arising from a detection of Kepler-13 C if the guess mass is correct. While we do not detect the spectral signature of Kepler-13 C for any guess mass from 0.2 - 1.5 $M_{\odot}$, we perform a sensitivity analysis by injecting synthetic companion spectra into the data and repeating the above procedure. The rate of detection is shown as a function of the effective temperature of the companion in Fig.\ \ref{Csensitivity}. 
This analysis indicates that we would detect a main sequence companion with effective temperature of $T_{\mathrm{eff}}>4700$ K (corresponding to a  mass $>0.75 M_{\odot}$) 95\% of the time, allowing us to set a mass limit on Kepler-13 C of $<0.75 M_{\odot}$ at 95\% confidence. Combined with the value of $M\sin i=0.4 M_{\odot}$ found by \cite{Santerne12}, we limit the mass of Kepler-13 C to $0.4 M_{\odot}<M<0.75 M_{\odot}$. We note that, as Kepler-13 C has not been directly detected, it could in principle be a white dwarf rather than a late-type dwarf (white dwarfs were not included in our spectral library for cross-correlation due to flux ratio issues). 

\section{Discussion}
\label{discussion}

Our best-fit value for the spin-orbit misalignment for Kepler-13 Ab, $\lambda=58.6^{\circ} \pm 2.0^{\circ}$, is in stark disagreement with the value of $\lambda=23^{\circ} \pm 4^{\circ}$ found by \cite{Barnes11}. Even if we fix $b$ to the value found by \cite{Barnes11}, we obtain a spin-orbit misalignment of $\lambda\sim54^{\circ}$, still in disagreement with the gravity darkening value. We do not have a definitive explanation for the mismatch between our result and that from \cite{Barnes11}. We note, however, that our value relies upon fewer assumptions regarding the physical nature of the star (e.g., the gravity-darkening law and gravity-darkening parameter), and thus is likely more robust. Additionally, \cite{Barnes11} fixed the effective temperature of the pole of Kepler-13 A to 8848 K, the temperature from the Kepler Input Catalog (KIC), rather than a more accurate spectroscopic value \citep[$T_{\mathrm{eff}}=8511^{+401}_{-383}$, from][though these values for the temperature differ by less than 1$\sigma$]{Szabo11}. The fact that Kepler-13 is a near-even flux ratio binary is also not accounted for in the KIC. \cite{Barnes11} could not account for any effects of the tertiary stellar companion Kepler-13 C upon the transit lightcurve, as this companion had not yet been discovered \citep{Santerne12}. Kepler-13 C, however, should contribute somewhere between 0.8\% and 0.03\% of the total flux of the system, given our limits on its mass, insufficient to significantly affect the dilution. Variability of either Kepler-13 B or C would need to occur on the orbital period of Kepler-13 Ab, or on a harmonic thereof, in order to systematically affect the light curve shape, which is unlikely. Finally, \cite{Barnes11} found a rotation period of 22.0 hours for Kepler-13 A by fitting their model to the data, slightly shorter than the likely rotation period of 25.4 hours found by \cite{Szabo14} in the {\it Kepler} data. While it is unclear whether the 25.4 hour period is indeed due to stellar rotation, if this is rotation then, given this and the likely too high value of $T_{eff}$ assumed by \cite{Barnes11}, the actual temperature (and therefore surface brightness) contrast between the poles and equator of Kepler-13 A should be smaller than that assumed by \cite{Barnes11}. The effects of this upon the lightcurve shape and the resulting inferred spin-orbit misalignment, however, are not qualitatively obvious, and a quantitative analysis is beyond the scope of this work.

As noted above, there is a great deal of disagreement in the literature as to the value of the impact parameter $b$, with published values ranging from 0.253 \citep{Szabo12} to 0.75 \citep{Mazeh12,Szabo11}. As noted earlier, these discrepancies cannot be attributed to precession of the planetary orbital plane \citep{Szabo12}. Our value of $b=0.256 \pm 0.030$ agrees to within $1\sigma$ only with the published measurement of \cite{Szabo12}, and is in disagreement with other published values by up to $16\sigma$. We note that our value of the impact parameter is obtained directly from the spectroscopy, and is thus largely independent of the previous measurements from the {\it Kepler} photometry (although our model requires the assumption of the transit duration from the photometry, as a prior in the MCMCs). This suggests a possible reason for the discrepancy between our value of $\lambda$ and that from \cite{Barnes11}. The value of $\lambda$ derived from gravity darkening is dependent upon the choice of impact parameter; as the value of $b=0.31962$ used by \cite{Barnes11} differs from the $b=0.256$ that we measure, it is perhaps unsurprising that the two values of $\lambda$ are in disagreement.

Using the value that \cite{Barnes11} measured for the stellar obliquity with respect to the line of sight \citep[$i=-45^{\circ}\pm4^{\circ}$; note that $i$ was denoted as $\psi$ by][]{Barnes11} and our measurement of $\lambda$, we calculate a full three-dimensional spin-orbit misalignment of $\varphi=73.5^{\circ} \pm 2.2^{\circ}$. Given the disagreement of our value of $\lambda$ with that from \cite{Barnes11}, however, it is unclear whether their measurement of $i$ remains applicable.

Despite the presence of an additional star in the Kepler-13 system, \cite{Barnes11} disfavor emplacement of Kepler-13 Ab via Kozai cycles due to the young system age \citep[$\sim700$ Myr, determined using isochrones by][]{Szabo11}, its current circular orbit \citep{Szabo11} or very small eccentricity \citep{Shporer14}, and the long timescale necessary for tidal semi-major axis damping. \cite{Barnes11} estimated that, for an initial Kozai-driven eccentric orbit similar to that currently occupied by HD 80606\,b, the required tidal damping timescale to circularize the orbit at Kepler-13 Ab's current location is $\sim 2\times10^{14}$ years. \cite{Barnes11} also noted that planet-planet scattering remains viable if it took place early enough that a debris disk sufficiently massive to quickly damp out the planetary eccentricity remained in place. Given the characteristics of the Kepler-13 and the highly inclined orbit that we find for Kepler-13 Ab, it seems natural that it could have been emplaced through migration within an inclined disk produced via the mechanism of \cite{Batygin12}. This would require an inclination between the orbital plane of Kepler-13 Ab and that of Kepler-13 BC about Kepler-13 A. Unfortunately, due to the lack of information about the position angle of Kepler-13 Ab's transit chord relative to the Kepler-13 AB separation, and the long orbital period of Kepler-13 BC about A (the projected separation is $\sim500$ AU), this relative inclination is unlikely to be measured in the foreseeable future. The mechanism proposed by \cite{Bate10} could also naturally result in an inclined, circular orbit for Kepler-13 Ab, but would not require the presence of a binary companion. We note, however, that these arguments rest upon the tidal circularization timescale being longer than the age of the system; as tidal theory continues to be not well understood, the eccentricity damping timescale may be very uncertain. Additionally, we note that due to these uncertainties we cannot definitively exclude any mechanisms for the emplacement of Kepler-13 Ab upon its current inclined orbit.

A 25.4-hour periodicity is evident in the {\it Kepler} lightcurves for Kepler-13. This was suggested to be either stellar pulsations \citep{Shporer11} or rotation \citep{Szabo12,Szabo14}. Additionally, \cite{Santerne12} found a 25.5-hour periodicity in their radial velocity measurements of Kepler-13 A. They noted that this radial velocity periodicity could also be due to either pulsations or rotation, but preferred the pulsation explanation because their measured radial velocity semi-amplitude of $1.41 \pm 0.38$ km s$^{-1}$ is much larger than that expected from starspots and rotation. We folded our stellar radial velocity for Kepler-13 Ab (i.e., the radial velocity offset between the solar and stellar barycentric rest frames discussed earlier) on the period found by \cite{Santerne12}, and our data appear to exhibit a similar periodicity and phase. In order to quantify this effect, we computed the generalized Lomb-Scargle periodograms \citep{ZechmeisterKurster09} for the \cite{Santerne12} dataset and our dataset. For the \cite{Santerne12} data we find a best-fitting period of 25.5 hours, and for our dataset, we find a period of 24.7 hours. The false alarm probabilities for these frequencies are 0.9998 and 0.98, respectively, and so we do not consider the detections of these periodicities in the radial velocity data to be statistically significant.

We see no evidence for stellar non-radial pulsations in our data, as are seen for the $\delta$ Sct planet host WASP-33 \citep[][and \S\ref{testing}]{CollierCameron10}, although given the short time span of each of our observations ($\sim1$ hour) such long-period pulsations would not necessarily manifest in our data. In principle we could compare the overall line shape for Kepler-13 A between different transit observations, but the moving line profile of Kepler-13 B would complicate such an effort, and thus we do not attempt such an analysis. We estimate that $\gamma$ Dor-like pulsations (similar in period to the Kepler-13 A periodicity, but typically exhibited by cooler stars)  would result in radial velocity shifts of order meters per second \citep[using the results of][]{Mathias04}, far too small to be detected in our data or to affect our conclusions.

The recently launched {\it Gaia} mission should be capable of further improving the characterization of the Kepler-13 system. {\it Gaia} is estimated to have an astrometric precision of $\sim5-14$ $\mu$as for stars with $6 < V < 12$ \citep{Eyer13}, like both Kepler-13 A and B. Thus, it should be capable of detecting both the mutual orbit of Kepler-13 A and BC ($\sim1$ mas yr$^{-1}$ for a circular, face-on orbit) and the orbit of Kepler-13 B about C (total displacement $\sim 200$ $\mu$as). Together with the radial velocity observations of \cite{Santerne12}, this will allow the measurement of the true mass of Kepler-13 C and its orbital plane. While the orbital period of Kepler-13 A around BC is likely too long to obtain a good orbital solution ($P\sim6000$ yr), {\it Gaia} should nonetheless be able to place some constraints upon the system parameters. The astrometric orbit of Kepler-13 A due to Kepler-13 Ab is too small to be detectable by {\it Gaia} (total displacement $\sim0.5$ $\mu$as).

\cite{Barnes11} note that, in principle, the spin-orbit misalignment for Kepler-13 Ab can be measured using a third mechanism: the photometric Rossiter-McLaughlin effect \citep{Shporer12,Groot12}. Unfortunately, given the scatter in the single-quarter \cite{Barnes11} lightcurve of $\sim40$ ppm, and that they estimate the amplitude of the photometric Rossiter-McLaughlin effect to be $\sim4$ ppm, this measurement is probably out of reach of even the full 16-plus quarter {\it Kepler} lightcurve.

\section{Conclusions}

We have constructed a Doppler tomography code which now rivals previously established codes in terms of precision. We have validated this code by analyzing data on a transit of WASP-33 b.
We have presented Doppler tomographic observations for the {\it Kepler} planet Kepler-13 Ab, finding a prograde orbit and measuring a much larger spin-orbit misalignment than that previously found by \cite{Barnes11} via the gravity darkened light curve. Given the disagreement between these two techniques, observations of further systems via both techniques will be of interest to determine the reason for the disagreement. 
We have also suggested that, due to its highly inclined, circular orbit, the (likely) long tidal damping timescale of the system, and the presence of a wide binary companion, Kepler-13 Ab may have been emplaced via migration within an inclined disk. Simulations of the system could confirm the viability of this hypothesis for the Kepler-13 system, but these are beyond the scope of the current work.

\vspace{12pt}

Thanks to Michael Endl, Edward L.\ Robinson, and Chris Sneden for guidance during this project; to Andrew Collier Cameron, for valuable discussions; to Douglas Gies and Zhao Guo, for consultations on velocity fields in the atmospheres of early-type stars; to Michel Breger, for consulting on the expected radial velocity pulsation amplitude of Kepler-13 A; and to the anonymous referee, for thorough comments that improved the quality of the paper.

M.C.J.\ gratefully acknowledges funding from a NASA Earth and Space Science Fellowship under Grant NNX12AL59H. This work was also supported by NASA Origins of Solar Systems Program grant NNX11AC34G to W.D.C. Work by S.A.\ was supported by NSF Grant No. 1108595. Work by S.D.R. was supported by NSF grant AST-1055910. J.N.W.\ was partly supported by the NASA Origins program and Kepler Participating Scientist program.

This paper includes data taken at The McDonald Observatory of The University of Texas at Austin. The Hobby-Eberly Telescope (HET) is a joint project of the University of Texas at Austin, the Pennsylvania State University, Stanford University, Ludwig-Maximilians-Universit\"at M\"unchen, and Georg-August-Universit\"at G\"ottingen. The HET is named in honor of its principal benefactors, William P. Hobby and Robert E. Eberly.

\clearpage
\begin{turnpage}
\begin{deluxetable}{lccccccccc}
\tabletypesize{\scriptsize}
\tablecolumns{10}
\tablewidth{0pt}
\tablecaption{Parameters of Kepler-13 Ab from the Literature \label{oldknowledge}}
\tablehead{
\colhead{Parameter} & \colhead{Placek et al.} & \colhead{M{\"u}ller et al.} & \colhead{Esteves et al.} & \colhead{Mazeh et al.} & \colhead{Mislis \& Hodgkin} & \colhead{Szab\'o et al.} & \colhead{Shporer et al.} & \colhead{Barnes et al.} & \colhead{Szab\'o et al.} \\
\colhead{} & \colhead{(2014)} & \colhead{(2013)} & \colhead{(2013)} & \colhead{(2012)} & \colhead{(2012)} & \colhead{(2012)} & \colhead{(2011)} & \colhead{(2011)} & \colhead{(2011)} 
}

\startdata
$R_p\ (R_J)$ & $>0.748$ & \ldots & $2.042$ & \ldots & \ldots & \ldots & \ldots & $1.445$ & $2.2$ \\
\ldots & $\pm0.015$ & \ldots & $\pm0.080$ & \ldots & \ldots & \ldots & \ldots & $\pm0.016$ & $\pm0.1$ \\
\hline
$M_p\ (M_J)$ & $8.35$ & \ldots & $7.95$ & $10$ & $8.3$ & $9.2$ & \ldots & \ldots & \ldots \\
\ldots & $\pm0.43$ & \ldots & $\pm0.27$ & $\pm2$ & $\pm1.25$ & $\pm1.1$ & \ldots & \ldots & \ldots \\
\hline
$P$ (days) & $1.76367$ & $1.763586522$ & $1.7635877$ & \ldots & \ldots & \ldots & $1.7637$ & \ldots & \ldots \\
\ldots & $\pm0.00007$ & $^{+0.000000194}_{-0.000000160}$ & $\pm0.000001$  & \ldots & \ldots & \ldots & $\pm0.0013$ & \ldots & \ldots \\
\hline
$b$ & \ldots & $0.323$ & $0.3681$ & $0.75$ & \ldots & $0.253$ & \ldots & 0.31598 & 0.75 \\
\ldots & \ldots & $^{+0.008}_{-0.007}$ & $^{+0.0041}_{-0.0064}$ & $\pm0.01$ &  \ldots & $\pm0.020$ & \ldots & \ldots & \ldots \\
\hline
$\lambda\ (^{\circ})$ & \ldots & \ldots & \ldots &\ldots & \ldots & \ldots & \ldots & $\pm23$ or $ \pm157$ & \ldots \\
\ldots & \ldots & \ldots & \ldots &\ldots & \ldots & \ldots & \ldots & $\pm4$ & \ldots \\
\hline
$i (^{\circ})$ & \ldots & \ldots & \ldots & \ldots & \ldots & \ldots & \ldots & $-48$ & \ldots \\
\ldots & \ldots & \ldots & \ldots & \ldots & \ldots & \ldots & \ldots & $\pm4$ & \ldots \\
\hline
$\varphi\ (^{\circ})$ & \ldots & \ldots & \ldots & \ldots & \ldots & \ldots & \ldots & $56$ & \ldots \\
\ldots & \ldots & \ldots & \ldots & \ldots & \ldots & \ldots & \ldots & $\pm4$ & \ldots \\
\hline
$i_P (^{\circ})$ & $81.37$ & $85.82$ & $85.135$ & \ldots & \ldots & \ldots & \ldots & $85.9$ & \ldots \\
\ldots & $\pm5.23$ & $^{+0.10}_{-0.12}$ & $^{+0.097}_{-0.063}$ & \ldots & \ldots & \ldots & \ldots & $\pm0.4$ & \ldots \\
\hline
$a/R_*$ & \ldots & $4.434$ & $4.3396$ & $3.17$ & \ldots & \ldots & \ldots & \ldots \\
\ldots & \ldots & $^{+0.011}_{-0.010}$ & $^{+0.0102}_{-0.0075}$ & $\pm0.06$ & \ldots & \ldots & \ldots & \ldots \\
\hline
$R_P/R_*$ & \ldots & $0.08553$ & $0.080509$ & $0.0907$ & \ldots & \ldots & \ldots & 0.084513 & $0.0884$ \\
\ldots & \ldots & $\pm0.000007$ & $^{+0.000033}_{-0.000048}$ & $\pm0.0005$ & \ldots & \ldots & \ldots & \ldots & $\pm0.0027$ \\
\hline
$P_{rot,*}$ (hr) & \ldots & \ldots & \ldots & \ldots & \ldots & $25.43$  & \ldots & 22.0 & \ldots \\
\ldots & \ldots & \ldots & \ldots & \ldots & \ldots & $\pm0.05$  & \ldots & \ldots & \ldots \\
\hline
$f_*$ & \ldots & \ldots & \ldots & \ldots & \ldots & \ldots & \ldots & 0.021 & \ldots \\
\enddata

\tablecomments{Values from \cite{Barnes11} assume a value of $M_*=2.05 M_{\odot}$, from \cite{Szabo11}. $b$ is the impact parameter, $\lambda$ is the projection of the spin-orbit misalignment onto the plane of the sky, $i$ is the stellar obliquity (denoted as $\psi$ by \cite{Barnes11}, $\pm{\varphi}$ is the full three-dimensional spin-orbit misalignment, $i_P$ is the inclination of the planetary orbit with respect to the plane of the sky (typically denoted $i$, but we adopt the notation $i_P$ to distinguish it from the stellar obliquity $i$), and $f_*=(R_{eq}-R_{pole})/R_{eq}$ is the stellar dynamical oblateness \citep{Barnes09}, where $R_{eq}$ and $R_{pole}$ are the stellar equatorial and polar radii, respectively.}

\end{deluxetable}

\end{turnpage}
\clearpage

\end{document}